\documentclass{article}
\usepackage{spconf,amsmath,graphicx}
\usepackage{amsfonts}
\usepackage[]{flushend}
\usepackage[hidelinks]{hyperref}
\usepackage{cite}
\usepackage[utf8]{inputenc}
\usepackage{pgfplots}
\DeclareUnicodeCharacter{2212}{−}
\usepgfplotslibrary{groupplots,dateplot}
\usetikzlibrary{patterns,shapes.arrows}
\pgfplotsset{compat=newest}

\usepackage{lipsum} %
\usepackage{tikzpagenodes} %

\usepackage{acronym}
\newacro{SNR}{signal-to-noise ratio}
\newacro{STFT}{short-time Fourier transform}
\newacro{STSA}{short-time spectral amplitude}
\newacro{MMSE}{minimum mean squared error}
\newacro{MSE}{mean squared error}
\newacro{TCN}{temporal convolution network}
\newacro{MAP}{\emph{maximum a posteriori}}
\newacro{A-MAP}{approximate MAP}
\newacro{ML}{maximum likelihood}
\newacro{iSTFT}{inverse STFT}
\newacro{Si-SDR}{scale-invariant signal-to-distortion ratio}
\newacro{DNS}{Deep Noise Suppression}
\newacro{ESTOI}{extended short-time objective intelligibility}
\newacro{PESQ}{perceptual evaluation of speech quality}
\newacro{DNN}{deep neural network}
\newacro{POLQA}{perceptual objective listening quality  analysis}
\newacro{NMF}{non-negative matrix factorization}
\newacro{VAE}{variational autoencoder}
\newacro{MCEM}{Monte Carlo expectation
maximization}
\newacro{WER}{word error rate}

\newcommand\mnoisy{\mathbf{X}_{ft}}
\newcommand\mspeech{\mathbf{S}_{ft}}
\newcommand\mego{\mathbf{E}_{ft}}

\newcommand\mnoise{\mathbf{N}_{ft}}

\newcommand\Left{\left(}
\newcommand\Right{\right)}
\newcommand\redone{\vspace{-0.1cm}}

\title{Partially adaptive Multichannel joint reduction of ego-noise \\and environmental noise}

\name{Huajian Fang$^{1,2}$\thanks{This work was funded by the DFG project number 261402652 and ahoi.digital.}, Niklas Wittmer$^{1}$, Johannes Twiefel$^{2,3}$,  Stefan Wermter$^{2}$, Timo Gerkmann$^1$}
\address{
  $^1$Signal Processing (SP), $^2$Knowledge Technology (WTM), Universität Hamburg, Germany\\
  $^3$exXxa GmbH, Hamburg, Germany
}

\usepackage[moderate,mathdisplays=tight]{savetrees}
\begin{document}
\ninept
\maketitle
\begin{abstract}
\vspace{-0.15cm}
Human-robot interaction relies on a noise-robust audio processing module capable of estimating target speech from audio recordings impacted by environmental noise, as well as self-induced noise, so-called ego-noise. While external ambient noise sources vary from environment to environment, ego-noise is mainly caused by the internal motors and joints of a robot. Ego-noise and environmental noise reduction are often decoupled, i.e., ego-noise reduction is performed without considering environmental noise. Recently, a variational autoencoder~(VAE)-based speech model has been combined with a fully adaptive non-negative matrix factorization~(NMF) noise model to recover clean speech under different environmental noise disturbances. However, its enhancement performance is limited in adverse acoustic scenarios involving, e.g.~ego-noise. In this paper, we propose a multichannel partially adaptive scheme to jointly model ego-noise and environmental noise utilizing the VAE-NMF framework, where we take advantage of spatially and spectrally structured characteristics of ego-noise by pre-training the ego-noise model, while retaining the ability to adapt to unknown environmental noise. Experimental results show that our proposed approach outperforms the methods based on a completely fixed scheme and a fully adaptive scheme when ego-noise and environmental noise are present simultaneously.
\end{abstract}

\vspace{-0.25cm}

\begin{tikzpicture}[remember picture,overlay]
  \node [draw=black, fill=white, text width=\textwidth, inner sep=7pt, yshift=-1.2cm] at (current page text area.south){\small{Accepted paper. \copyright~2023 IEEE. Personal use of this material is permitted. Permission from IEEE must be obtained for all other uses, in any current or future media, including reprinting/republishing this material for advertising or promotional purposes, creating new collective works, for resale or redistribution to servers or lists, or reuse of any copyrighted component of this work in other works.}};
\end{tikzpicture}

\section{Introduction}
\label{sec:intro}
\vspace{-0.25cm}
In recent decades, research on autonomous systems (AS) such as humanoid interactive robots has received increasing attention. Interactive robots are typically equipped with multiple microphones to perceive their environment and react to requests or particular commands from humans. However, the acquisition of target acoustic information is often disturbed not only by external interfering sources, i.e., environmental noise, but also by self-generated noise, also called \emph{ego-noise}. It poses difficulties for subsequent tasks, such as speech recognition and language understanding. This calls for a noise-robust audio processing module capable of recovering target clean speech to support the robot's actuator unit to act appropriately~\cite{schmidt2020acoustic, stefan2019asr}.

In human-robot interaction, ego-noise may originate from different parts of the robot and reducing ego-noise is non-trivial in various aspects. It is mainly caused by the electric motors and mechanical parts distributed all over the robot body~\cite{deleforge2019audio,deleforge2015ksvd}. The microphones are often placed close to the motors, especially for small-sized robots, resulting in acoustic scenarios with challenging \acp{SNR}. %
Furthermore, as ego-noise coming from, e.g.~robotic limb movements, is non-stationary, it may be considered a difficult noise source. However, due to the limited degree of motion, ego-noise from the motors and joints exhibits a characteristic spatial and spectral structure. Thus, specialized and efficient light-weight machine learning algorithms can be designed to learn and exploit these distinct spatial and spectral characteristics of ego-noise~\cite{schmidt2020acoustic,deleforge2019audio,deleforge2015ksvd,alexander2018motor,schmidt2021mnmfdata,haubner2018multichannel,ito2005internal,tezuka2014ego,ince2011assessment}. 

For instance, ego-noise can be modeled by dictionary-based algorithms, e.g.~\ac{NMF}~\cite{lee2000algorithms,tezuka2014ego, schmidt2020motor}, where ego-noise is approximated by a linear combination of pre-captured dictionary components. For multichannel recordings, in addition to structured tempo-spectral characteristics, spatial information can also be employed using, e.g.~multichannel \ac{NMF}~\cite{schmidt2021mnmfdata,haubner2018multichannel}. Deleforge et al.~\cite{deleforge2015ksvd} have proposed a sparse representation of multichannel ego-noise signals in the complex domain. Some approaches have included information from other modalities, such as motor data~\cite{alexander2018motor,ito2005internal, schmidt2020motor}. However, this requires synchronized multimodal data, which may not be readily available. While a pre-learned ego-noise model has shown some effectiveness in modeling noise characteristics, it may cause noise mismatch problems in realistic scenarios that include not only ego-noise, but also unknown environmental noise signals. 

Currently, advanced methods for environmental noise reduction are based on deep neural networks (DNNs)~\cite{wang2018supervised}. The \ac{VAE} is a deep generative model that can be used to learn a probabilistic prior distribution of clean speech~\cite{kingma2019introduction}. It has been combined with a statistical \ac{NMF} noise model to perform speech enhancement, where the \ac{VAE}-based speech model is pre-trained on clean speech while the parameters of the \ac{NMF} model are estimated based on noisy observations~\cite{simon2018vae,kouhei2018mvae,simon2019vaemnmf}. The \ac{VAE}-\ac{NMF} framework has shown improved speech enhancement performance and generalization capabilities over its \ac{NMF} counterpart and fully supervised baselines~\cite{simon2018vae,kouhei2018mvae,simon2019vaemnmf}. While the fully adaptive \ac{NMF} noise model can potentially adapt to various acoustic scenarios, gaining robustness under adverse acoustic conditions (e.g.~when ego-noise and environmental noise are present simultaneously) remains a challenging task, as we will show in experiments. Few existing publications take both ego-noise and environmental noise into account~\cite{ince2011assessment, fang2021partial,ince2011audition}. Ince et al. proposed to reduce stationary background noise independently of ego-noise~\cite{ince2011assessment}. Our previous work~\cite{fang2021partial} has presented a single-channel joint noise reduction system for interactive robots, but disregarded spatial information.

In this work, we propose a multichannel joint ego-noise and environmental noise reduction method for interactive robots. For this, the tempo-spectral features of speech are modeled using the \ac{VAE} and the noise characteristics are modeled by multichannel NMF as in~\cite{simon2019vaemnmf}. More specifically, similar to multichannel ego-noise approaches such as \cite{haubner2018multichannel}, we want to take advantage of spatially and spectrally structured characteristics of ego-noise to gain robustness in adverse conditions. At the same time, similar to, e.g.~\cite{simon2019vaemnmf}, we want to retain the adaptation ability to unknown environmental noise. For this, we propose to model ego-noise and environmental noise separately. We pre-train the ego-noise model to capture the spectral and spatial features, while its temporal activation is adapted to noisy observations jointly with the parameters of the environmental noise model. Experimental results show the considerable benefits of the proposed joint reduction method when ego-noise and environmental noise are present simultaneously.

\vspace{-0.25cm}
\section{Signal Model}
\label{sec:signalmodel}
\vspace{-0.2cm}
We consider an acoustic scenario where the target speech signal is disturbed by additive noise and recorded by a microphone array with $M$ channels. We transform the noisy mixture into the time-frequency domain using the~\ac{STFT}:
\begin{equation}
    \mnoisy = \sqrt{g_t}\mspeech + \mnoise \, ,
\end{equation}
where $\mnoisy\in\mathbb{C}^M$, $\mspeech \in \mathbb{C}^M$, and $\mnoise \in \mathbb{C}^M$ represent the complex coefficients of the mixture signal, the speech signal, and the noise signal at the frequency bin $f\in\{1,\cdots,F\}$ and the time frame $t\in \{1,\cdots,T\}$. $g_t$ is a gain parameter to increase the robustness to the time-varying loudness of speech sounds~\cite{simon2019vaemnmf}. Note that the noise signals $\mnoise$ may contain either ego-noise or environmental noise or both. We aim to recover clean speech with improved quality and intelligibility given only noisy mixtures. 

\vspace{-0.15cm}
\subsection{Noise model}
\vspace{-0.1cm}
The noise coefficients are assumed to follow a complex Gaussian
distribution with zero mean
\begin{equation}
\redone
    \mnoise \sim \mathcal{N}_\mathbb{C}(\boldsymbol{0},\ \mathbf{\Sigma}_{N,ft}) \, ,
\redone
\end{equation}
where $\mathcal{N}_{\mathbb{C}}(\boldsymbol{\mu}, \mathbf{\Sigma})$ denotes the complex Gaussian distribution with mean $\boldsymbol{\mu}$ and covariance matrix $\mathbf{\Sigma}$. The covariance matrix is defined as
\begin{equation}
\label{eq:noisespatialmodel}
\mathbf{\Sigma}_{N,ft} = \mathbf{R}_{N,f}\sigma_{N,ft}^2,
\end{equation}
where $\mathbf{R}_{N,f} \in \mathbb{C}^{M\times M}$ is a spatial covariance matrix characterizing the sound propagation process from sources to microphones. $\sigma_{N,ft}^2$ represents the noise spectral variance, which can be modeled using the \ac{NMF},
\begin{equation}
\label{eq:egonmf}
    \redone
    \sigma_{N,ft}^2 = \left[\mathbf{W}_N\mathbf{H}_N\right]_{ft}=\sum_{k=1}^Kw_{fk}h_{kt}
    \, , 
\redone
\end{equation}
where $\mathbf{W}_N\in \mathbb{R}^{F\times K}_{+}$ denotes the dictionary matrix that captures the time-frequency characteristics of noise and $\mathbf{H}_N\in \mathbb{R}^{K\times T}_{+}$
denotes the coefficient matrix that represents the temporal activity. The noise dictionary contains $K$ atoms indexed by $k$ ($K$ is also referred to here as the dictionary size). We will decompose the noise signal $\mathbf{N}_{ft} = \mathbf{E}_{ft} + \mathbf{B}_{ft}$ into ego-noise $\mathbf{E}_{ft}$ and environmental noise $\mathbf{B}_{ft}$ in Section~\ref{sec:jointreduction}.
\vspace{-0.15cm}
\subsection{Speech model}
\vspace{-0.1cm}
We assume that the clean speech coefficients are complex Gaussian-distributed:
\begin{equation}
    \mspeech \sim \mathcal{N}_\mathbb{C}(\boldsymbol{0},\ \mathbf{\Sigma}_{S,f}\left(\mathbf{z}_t)\right) \, ,
\end{equation}
where $\mathbf{\Sigma}_{S,f}(\mathbf{z}_t) = \mathbf{R}_{S,f}\sigma_{S,f}^2(\mathbf{z}_t)$. $\mathbf{R}_{S,f}\in \mathbb{C}^{M\times M}$ is the speech spatial covariance matrix. It is assumed that the speech tempo-spectral power can be inferred from the latent variable $\mathbf{z}_t\in \mathbb{R}^L$, denoted as $\sigma_{S,f}^2(\mathbf{z}_t)$, which can be realized by the generative model of the \ac{VAE}, i.e., the decoder. Let $\mathbf{s}_t \in \mathbb{C}^F$ be a vector of single-channel clean speech spectra at the $t$-th time frame. The posterior of the latent variable $q(\mathbf{z}_t|\mathbf{s}_t)$ is approximated by a real-valued Gaussian distribution 
\begin{equation}
\mathbf{z}_t|\mathbf{s}_t \sim  \mathcal{N}(\mu_z(|\mathbf{s}_t|^2),\ \sigma_z(|\mathbf{s}_t|^2))\, ,
\end{equation}
where $\mu_z(|\mathbf{s}_t|^2): \mathbb{R}_+^F \rightarrow \mathbb{R}^L$ and $\sigma_z(|\mathbf{s}_t|^2): \mathbb{R}_+^F \rightarrow \mathbb{R}^L_+$ denote the nonlinear mapping from the power spectrogram to the mean and variance of the latent variable, implemented by the encoder of the \ac{VAE}, also called the recognition model. The parameters of the VAE can be jointly learned by maximizing the variational lower bound of the log-likelihood $\log p(\mathbf{s}_t)$
\begin{equation}
    \label{eqn:elbo}
    \mathcal{L}_{\text{VAE}} =
     \mathbb{E}_{q(\mathbf{z}_t|\mathbf{s}_t)}
     \left[\log p(\mathbf{s}_t|\mathbf{z}_t)\right] - \mathbb{KL}(q(\mathbf{z}_t|\mathbf{s}_t)||p(\mathbf{z}_t)) \, ,
\end{equation}
where $\mathbb{KL}(\cdot|\cdot)$ denotes the Kullback-Leibler divergence and $p(\mathbf{z}_t)$ represents the standard Gaussian prior of $\mathbf{z}_t$.

\vspace{-0.15cm}
\subsection{Clean speech estimation}
\vspace{-0.05cm}
With the assumption that the speech and noise signals are independent, the noisy mixture is given by 
\begin{equation}
    \mnoisy \sim \mathcal{N}_{\mathbb{C}}(\mathbf{0},\ g_t\mathbf{\Sigma}_{S,f}(\mathbf{z}_t)+\mathbf{\Sigma}_{N,ft}).
    \label{eq:previousmixture}
\end{equation}
The parameters of the VAE-based speech model are obtained by training the neural network on clean speech data. At testing, a \ac{MCEM} method can be employed to estimate the unknown parameters~\ \{$\mathbf{W}_N, \mathbf{H}_N, \mathbf{R}_{N,f}, \mathbf{R}_{S,f}, g_t$\}~\cite{simon2019vaemnmf}. Finally, the multichannel Wiener filter is employed to extract clean speech 
\begin{equation}
\redone
    \widehat{\mathbf{S}}_{ft} = g_t\mathbf{\Sigma}_{S,f}(\mathbf{z}_t)\left(g_t\mathbf{\Sigma}_{S,f}(\mathbf{z}_t) + \mathbf{\Sigma}_{N,ft}\right)^{-1}\mnoisy.
\end{equation}
The fully adaptive scheme in~\cite{simon2019vaemnmf} that optimizes the unknown parameters based on noisy inputs, will adapt flexibly to different types of noise without the need of prior information on potential noise structures. The main idea of this approach is to achieve a high generalization ability and robustness to unexpected noise types. However, if accurate prior knowledge is available, it can be very helpful to improve robustness, especially in acoustically challenging environments. Therefore, as ego-noise exhibits a very distinct spatial-spectral structure, prior knowledge can be efficiently exploited by pre-learning the dictionary matrix and the spatial covariance matrix on ego-noise recordings only. However, when only pre-learned on ego-noise, the flexibility and generalization to unseen scenarios is lost. For instance, rather poor performance is to be expected in environmental noise, which limits its applicability in realistic scenarios that contain both environmental noise and background noise.

\section{Joint reduction of ego-noise \\and environmental noise}
\label{sec:jointreduction}
\vspace{-0.15cm}
In this section, we present a multichannel partially adaptive scheme, where we improve noise modeling capabilities by decomposing noise into ego-noise and environmental noise. This allows us to obtain a robust prior pre-learned on the distinct spatial and spectral characteristics of ego-noise, while retaining the flexibility to adapt to environmental noise signals. 

\vspace{-0.15cm}
\subsection{Mixture model and speech estimation}
In a real-world human-robot interaction scenario, a target speech signal may be distorted by ego-noise and environmental noise simultaneously. We, thus, consider a noise model that is comprised of ego-noise $\mathbf{E}_{ft}$  and environmental noise $\mathbf{B}_{ft}$ as follows:
\begin{equation}
    \mathbf{N}_{ft} = \mathbf{E}_{ft} + \mathbf{B}_{ft}\,.
\end{equation}
By assuming that the ego-noise, environmental noise and speech signals are independent and complex Gaussian distributed, the noisy mixture follows a complex Gaussian of the form:
\begin{equation}
\label{eq:jointnoisymodel}
    \mnoisy \sim \mathcal{N}_{\mathbb{C}}(\mathbf{0},\ g_t\mathbf{\Sigma}_{S,f}(\mathbf{z}_t)+\mathbf{\Sigma}_{E,ft}+\mathbf{\Sigma}_{B,ft})\, ,
\end{equation}
where the covariance matrix  of environmental noise is defined as $\mathbf{\Sigma}_{B,ft}= \mathbf{R}_{B,f}\left[\mathbf{W}_B\mathbf{H}_B\right]_{ft}$ with $\left[\mathbf{W}_B\mathbf{H}_B\right]_{ft}=\sum_{k_b=1}^{K_B}w_{fk_b}h_{k_bt}$, and the covariance matrix of ego-noise as $\mathbf{\Sigma}_{E,ft} = \mathbf{R}_{E,f}\left[\mathbf{W}_E\mathbf{H}_E\right]_{ft}$ with $\left[\mathbf{W}_E\mathbf{H}_E\right]_{ft}=\sum_{k_e=1}^{K_E}w_{fk_e}h_{k_et}$. $K_B$ and $K_E$ are the sizes of the environmental noise dictionary and the ego-noise dictionary, respectively.

Similarly, clean speech can be estimated by applying the multichannel Wiener filter
\begin{equation}
    \widehat{\mathbf{S}}_{ft} = g_t\mathbf{\Sigma}_{S,f}(\mathbf{z}_t)(\mathbf{\Sigma}_{X,ft}(\mathbf{z}_t))^{-1}\mnoisy,
\end{equation}
where $\mathbf{\Sigma}_{X,ft}(\mathbf{z}_t)= g_t\mathbf{\Sigma}_{S,f}(\mathbf{z}_t)+\mathbf{\Sigma}_{E,ft}+\mathbf{\Sigma}_{B,ft}$. This requires estimating the unknown parameters~$\{\mathbf{W}_E,\mathbf{H}_E,\mathbf{W}_B,\mathbf{H}_B,\mathbf{R}_{S,f},\mathbf{R}_{E,f},\mathbf{R}_{B,f}, \\ g_t\}$. The following subsections describe the estimation of the ego-noise dictionary matrix $\mathbf{W}_{E}$ and the spatial covariance matrix~$\mathbf{R}_{E,f}$ using the pre-training technique, and an \ac{MCEM} optimization method to the proposed partially adaptive scheme. 
\vspace{-0.2cm}
\subsection{Training phase}
\vspace{-0.2cm}
To capture the spectral and spatial characteristics of ego-noise, we train a multichannel \ac{NMF} model on ego-noise recordings by optimizing the negative log-likelihood:
\begin{equation}
        \mathcal{L} = \sum_{f=1,t=1}^{F,T}\text{tr}\Left\mego\mego^H\mathbf{\Sigma}_{E,ft}^{-1}\Right +\ln \; \text{det}\Left\mathbf{\Sigma}_{E,ft}\Right,
\end{equation}
where constant terms are omitted~\cite{sawada2012mnmf}. $\text{tr}(\cdot)$ denotes the trace operator; $\text{det}(\cdot)$ denotes the determinant of a matrix; $\cdot^H$ denotes the conjugate transpose. Minimizing this function using the majorization scheme leads to the multiplicative update rules for $\{\mathbf{W}_E,\mathbf{H}_E,\mathbf{R}_{E,f}\}$~\cite{simon2019vaemnmf,sawada2012mnmf}. We fix the dictionary matrix $\mathbf{W}_E$ and the spatial covariance matrix $\mathbf{R}_{E,f}$ at the testing phase, while keeping $\mathbf{H}_E$ adaptive to noisy observations to account for different temporal variations.

\vspace{-0.2cm}
\subsection{Parameter optimization}
\label{sec:parameteroptimization}
\vspace{-0.1cm}
To estimate the unknown parameters in the testing phase, we follow the \ac{MCEM} optimization scheme by~\cite{simon2019vaemnmf}. At the \emph{Expectation step}, the complete-data log-likelihood is approximated by averaging over $R$ samples:
\begin{equation}
\begin{split}
    \mathbf{Q}(\theta;\theta^\ast) &= \mathbb{E}_{p(\mathbf{z}|\mathbf{X};\theta)}\left[\ln p(\mathbf{X},\mathbf{z};\theta)\right]\\
    &\approx -\frac{1}{R}\sum_{r=1}^R\sum_{f=1,t=1}^{F,T}\biggl[\text{tr}\left(\mnoisy\mnoisy^H\left[\mathbf{\Sigma}_{X,ft}\Left\mathbf{z}_t^{(r)}\Right\right]^{-1}\right) \\
    &+\ln \; \text{det}\left(\mathbf{\Sigma}_{X,ft}\Left\mathbf{z}_t^{(r)}\Right\right)\biggr]\,.
\end{split}
\label{eq:likelihood}
\end{equation}
$R$ samples of the latent variable are drawn using the Metropolis-Hastings algorithm with a Gaussian as a symmetric proposal distribution. $\theta^\ast$ is an initialization of the parameters.
At the \emph{Maximization step}, we minimize the loss function, i.e., the negative log-likelihood $-R\mathbf{Q}(\theta;\theta^\ast)$, with respect to the unknown parameters~$\theta=\{\mathbf{H}_E, \mathbf{W}_B,\mathbf{H}_B, \mathbf{R}_{S,f},\mathbf{R}_{B,f}, g_t\}$ using the auxiliary function technique. For this, equation~(\ref{eq:likelihood}) can be viewed as the superposition of a convex function~(the first term) and a concave function~(the second term), where the former can be bounded using the Jensen's trace inequality and the latter can be bounded using a first-order Taylor expansion~\cite[Appendix A]{simon2019vaemnmf}. This gives an upper bound function and computing the partial derivative with respect to each parameter separately leads to the iterative update rules:
\vspace{-0.2cm}
\begin{equation}
    g_t = g_t^* \left[
\frac{ \displaystyle
\sum_{r=1}^{R} \sum^{F}_{f=1}
\sigma^2_f \left( \mathbf{z}_t^{(r)} \right)\text{tr}\left[
\mathbf{M}_{ft}^{(r)} \mathbf{R}_{S,f}\right]
} { \displaystyle
\sum^{R}_{r=1} \sum^{F}_{f=1} \sigma^2_f \left( \mathbf{z}_t^{(r)} \right)
\text{tr}\left[\Left
\mathbf{\Sigma}_{X,ft}\Left\mathbf{z}_t^{(r)}\Right\Right^{-1} \mathbf{R}_{S,f}\right]
}
\right]^{\frac{1}{2}} ,
\end{equation}
\vspace{-0.2cm}
\begin{equation}
    w_{fk_b} = w_{fk_b}^* \left[
\frac{ \displaystyle
\sum_{r=1}^{R} \sum^{T}_{t=1}
h_{k_bt}\text{tr}\left[
\mathbf{M}_{ft}^{(r)} \mathbf{R}_{B,f}\right]
} { \displaystyle
\sum^{R}_{r=1} \sum^{T}_{t=1} h_{k_bt}
\text{tr}\left[\Left
\mathbf{\Sigma}_{X,ft}\Left\mathbf{z}_t^{(r)}\Right\Right^{-1} \mathbf{R}_{B,f}\right]
}
\right]^{\frac{1}{2}} ,
\end{equation}
\vspace{-0.2cm}
\begin{equation}
    h_{k_bt} = h_{k_bt}^* \left[
\frac{ \displaystyle
\sum_{r=1}^{R} \sum^{F}_{f=1}
w_{fk_b}\text{tr}\left[
\mathbf{M}_{ft}^{(r)} \mathbf{R}_{B,f}\right]
} { \displaystyle
\sum^{R}_{r=1} \sum^{F}_{f=1} w_{fk_b}
\text{tr}\left[\Left
\mathbf{\Sigma}_{X,ft}\Left\mathbf{z}_t^{(r)}\Right\Right^{-1} \mathbf{R}_{B,f}\right]
}
\right]^{\frac{1}{2}} ,
\end{equation}
\vspace{-0.2cm}
\begin{equation}
    h_{k_et} = h_{k_et}^* \left[
\frac{ \displaystyle
\sum_{r=1}^{R} \sum^{F}_{f=1}
w_{fk_e}\text{tr}\left[
\mathbf{M}_{ft}^{(r)} \mathbf{R}_{E,f}\right]
} { \displaystyle
\sum^{R}_{r=1} \sum^{F}_{f=1} w_{fk_e}
\text{tr}\left[\Left
\mathbf{\Sigma}_{X,ft}\Left\mathbf{z}_t^{(r)}\Right\Right^{-1} \mathbf{R}_{E,f}\right]
}
\right]^{\frac{1}{2}} ,
\end{equation}
\vspace{-0.2cm}
where $\mathbf{M}_{ft}^{(r)} = 
\Left\mathbf{\Sigma}_{X,ft}\Left\mathbf{z}_t^{(r)}\Right\Right^{-1}\mathbf{X}_{ft}\mathbf{X}_{ft}^{H} \Left\mathbf{\Sigma}_{X,ft}\Left\mathbf{z}_t^{(r)}\Right\Right^{-1}$.

\vspace{0.2cm}
The two adaptive spatial covariance matrices $\mathbf{R}_{S,f}$, $\mathbf{R}_{B,f}$ are updated by solving the corresponding algebraic Riccati equations as in the fully adaptive scheme~\cite{simon2019vaemnmf}~\cite[Appendix I]{sawada2012mnmf}.

\begin{figure}[t!]
\centerline{\includegraphics[width=6cm]{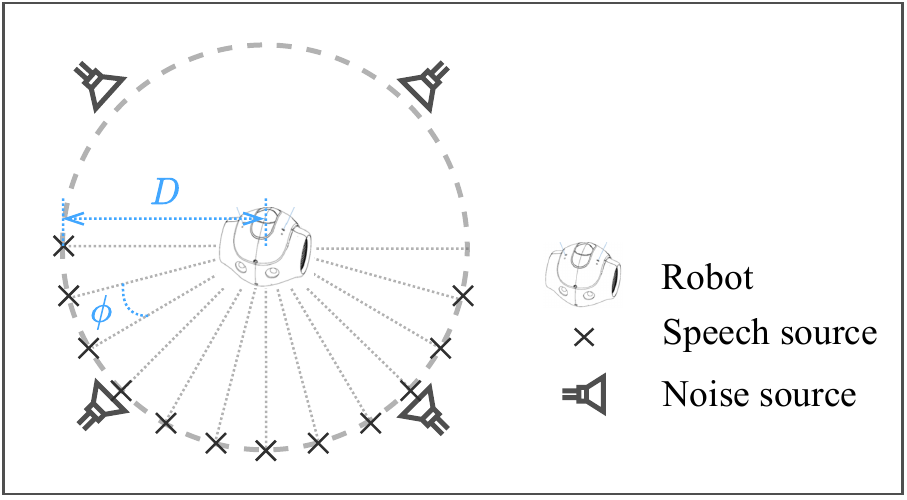}}
\vspace{-0.15cm}
\caption{Illustration of the recording setup with the NAO robot~\cite{aldebarannao}. Room dimensions~($\text{length}\times \text{width}\times\text{height}$): $504 \times 930 \times 284$~cm; $T_{60}\approx200~\text{ms}$; speaker-robot distance $D\approx1~\text{m}$; $\phi\approx15^{\circ}$.}
\label{fig:recording} 
\vspace{-0.55cm}
\end{figure}

\vspace{-0.1cm}
\section{Experiments}
\label{sec:experiment}
\vspace{-0.1cm}
In this section, the proposed partially adaptive scheme (referred to as \emph{Partial}) is compared to two baselines: 
\begin{itemize}
    \item \emph{Adaptive}: Refers to the fully adaptive scheme with all unknown parameters estimated based on noisy observations~\cite{simon2019vaemnmf}.
    \item \emph{Fixed}: Refers to the fixed scheme with the dictionary matrix and spatial covariance matrix pre-learned on ego-noise recordings at training time and fixed at test time as in the ego-noise reduction literature, e.g.~\cite{haubner2018multichannel,schmidt2021mnmfdata}.
\end{itemize}
For each adaptive scheme, 7 different dictionary sizes are considered, leading to a total of 21 compared methods. We evaluate the algorithms in two application scenarios:
\begin{itemize}
    \item \emph{Ego}: Only ego-noise is present, mimicking a scene where a person is talking to a robot performing certain movements.
    \item \emph{Ego + Env}: In addition to ego-noise, environmental noise is present simultaneously as an additional disturbance. 
\end{itemize}

We use the \ac{Si-SDR} measured in dB to account for both noise reduction and the speech artifacts~\cite{le2019sdr}, and the \ac{POLQA} to measure speech quality~\cite{polqa}. The speech recognition accuracy is measured by the~\ac{WER}. We employ the pre-trained speech recognition model Quartznet~\cite{kriman2020asr} in the NeMo toolkit~\cite{kuchaiev2019nemo}, in conjunction with a 4-gram language model available via the LibriSpeech website~\cite{librilr}.

\vspace{-0.2cm}
\subsection{Dataset}
\label{dataset}
\vspace{-0.1cm}
All algorithms are trained and evaluated on a dataset recorded in our varechoic chamber. We use a humanoid interactive robot NAO H25 from Softbank for recording purposes~\cite{gouaillier2009mechatronic}. The clean speech utterances are randomly chosen from the TIMIT test set~\cite{timit1993}. Each target clean speech sample is played through a loudspeaker randomly placed among the positions shown in Fig.~\ref{fig:recording} and recorded using external omnidirectional electret microphones mounted in the same position of the built-in microphone array~($M=4$) on the robot.  Ego-noise is recorded when the robot performs pre-defined right-arm movements in a crouching posture. To simulate external environmental noise sources, we re-record audio samples randomly selected from the DEMAND database~\cite{thiemann2013diverse} and the loudspeaker emitting environmental noise is placed at one of the four positions shown in Fig.~\ref{fig:recording}. For the ego-noise only scenario, we mix speech signals with out-of-training ego-noise recordings (with movement speeds different from training data) at \acp{SNR} randomly chosen from \{-5~dB, -4~dB, $\cdots$, 5~dB\}. To simulate a challenging realistic scenario, besides ego-noise, we further corrupt speech signals by environmental noise at a \ac{SNR} of 0~dB~\cite{simon2019vaemnmf}. In total, this leads to a test set of 128 noisy samples for each evaluation scenario, with an average \ac{SNR} of ${-2.1}$~dB for the joint noise scenario and ${-1.8}$~dB for the ego-noise only scenario.
\redone
\begin{figure}[t]
  \centerline{\includegraphics[width=8.4cm, height=6cm]{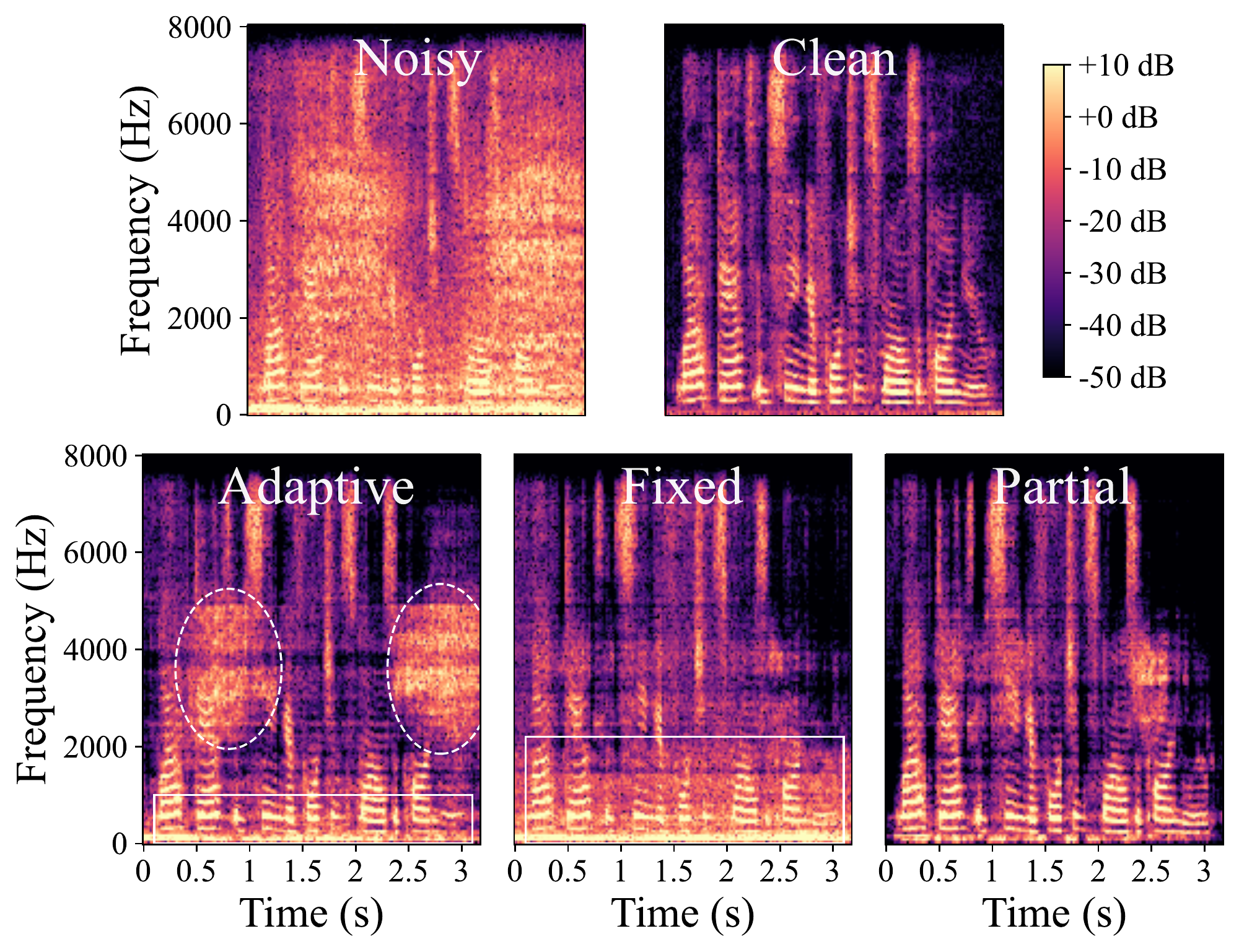}}
  \vspace{-0.3cm}
\caption{Spectrograms of an audio example. Clean speech is distorted by both ego-noise and environmental noise. The three plots in the second row represent the reconstructed speech spectrograms obtained by three compared methods.}
\label{fig:spectrogram} 
\vspace{-0.4cm}
\end{figure}
\redone
\vspace{-0.1cm}
\subsection{Hyperparameter settings}
\vspace{-0.1cm}
We use an \ac{STFT} with a Hann window of 64~ms and a hop size of 25~\%. All audio signals are sampled at 16~kHz. The decoder of the \ac{VAE} has two hidden layers of sizes 128 and 512 respectively. The hyperbolic tangent activation function is applied to the hidden layers; the linear activation is applied to the output layer. The encoder network consists of two hidden layers of sizes 512 and 128, respectively, with the hyperbolic tangent activation functions applied. The latent dimension $L$ is set to 16. The \ac{VAE} is trained on the re-recorded TIMIT training set using the same microphone setup as described in Section~\ref{dataset}. The network parameters are optimized using the Adam optimizer with a learning rate of 0.001 and a patience of 5 epochs. The parameters of the \ac{MCEM} algorithm are set as in~\cite{simon2019vaemnmf}, i.e., $R=10$ with a burn-in phase 30 iterations. For the partially adaptive scheme, we set the dictionary sizes for the fixed and adaptive parts as shown in Table~\ref{dictioanryparameters}.
\begin{table}[ht]
\centering
\vspace{-0.1cm}
\begin{tabular}{c|c|c|c|c|c|c|c}
 \hline
Total dictionary size
 & 16 & 32&64&96&128&160&192   \\
 \hline
 $K_B$ &8 &16&32&32&32&32&32   \\
 $K_E$ & 8 & 16&32&64&96&128&160   \\
 \hline
\end{tabular}
  \vspace{-0.1cm}
  \caption{Dictionary sizes for the proposed partially adaptive scheme.}
  \label{dictioanryparameters}
  \vspace{-0.5cm}
\end{table}

\begin{figure}[th]
  \centerline{\includegraphics[width=8.6cm,height=10.7cm]{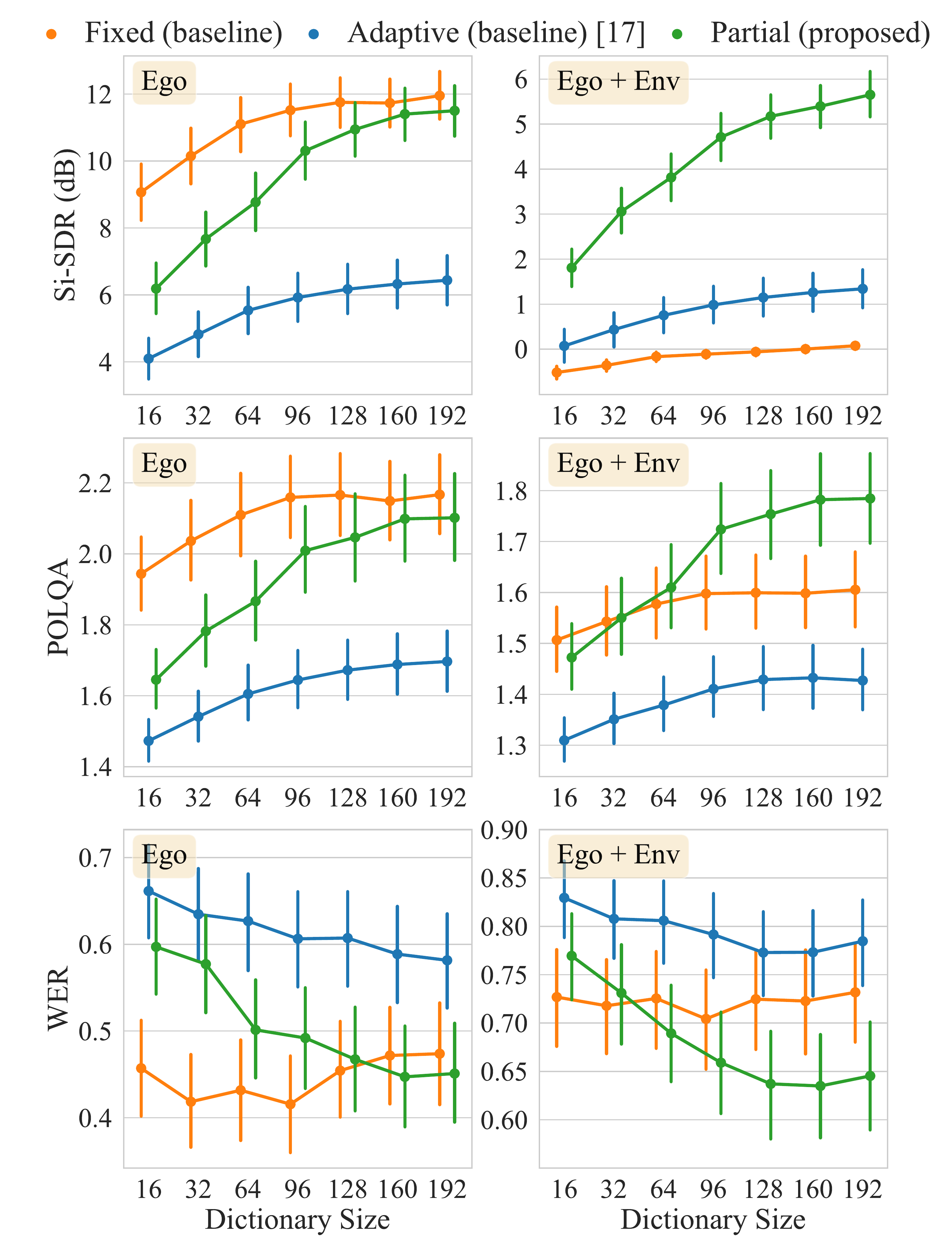}}
  \vspace{-0.3cm}
\caption{Higher  \ac{Si-SDR} and POLQA scores indicate better enhancement performance, and lower \ac{WER} indicates higher recognition accuracy. The marker denotes the mean value and the vertical bar indicates the 95\%-confidence interval.}
\vspace{-0.4cm}
\label{fig:metrics} 
\end{figure}
\vspace{-0.1cm}
\subsection{Results}
\vspace{-0.1cm}
The benefits of the partially adaptive scheme are visible in Fig.~\ref{fig:spectrogram}. While the fully adaptive scheme possesses the flexibility to adapt to various noisy conditions, its ability in capturing noise characteristics is limited especially when both ego-noise and environmental noise are present. This is shown by the residual ego-noise marked with the dashed ellipses and residual environmental noise marked with the solid rectangle in the reconstructed speech spectrogram. While the fixed scheme, whose reconstructed spectrogram is visualized in the second plot in the second row of Fig.~\ref{fig:spectrogram}, shows some effectiveness in removing ego-noise, the residual environmental noise is still quite pronounced, as marked by the solid rectangle. Finally, it can be observed that the proposed partially adaptive scheme shows a higher noise reduction effect than the other two approaches.

The two columns in Fig.~\ref{fig:metrics} display the evaluation results for the ego-noise only scenario~(\emph{Ego}) and the joint noise scenario~(\emph{$\text{Ego}+\text{Env}$}), respectively. We observe that the fully adaptive approach is outperformed by the fully fixed and partially adaptive schemes in the presence of ego-noise only, as shown by its lowest \ac{POLQA} and \ac{Si-SDR} scores and the highest \ac{WER}. This again implies that the fully adaptive scheme has difficulty in capturing ego-noise characteristics. The partially adaptive scheme and the fully adaptive scheme perform comparably when we increase the total dictionary size, indicating that ego-noise can be better modeled with a larger dictionary size due to its complexity and broadband characteristics. Eventually, it can be observed that the partially adaptive scheme delivers superior results over the other two methods when both noise types are present simultaneously. This indicates that with an appropriate dictionary size, the partially adaptive scheme can effectively approximate ego-noise while properly capturing unknown environmental noise in adverse scenarios. Audio examples are
available online\footnote{\url{https://uhh.de/inf-sp-mcpartial2023
}}.

\vspace{-0.2cm}
\section{Conclusion}
\label{sec:conclusion}
\vspace{-0.2cm}
Based on the deep generative model and multichannel \ac{NMF}, we proposed to jointly model ego-noise and environmental noise with a partially adaptive scheme. To exploit the spectrally and spatially structured characteristics of ego-noise, we pre-train the ego-noise model while keeping the environmental noise model adaptive to noisy observations. The proposed partially adaptive scheme demonstrated an increased performance compared to the approaches based on the fixed scheme and on the fully adaptive scheme in adverse scenarios where both ego-noise and environmental noise are present. %

\bibliographystyle{IEEEbib}
\bibliography{strings,refs}

\end{document}